\documentclass[preprintnumbers,amsmath,amssymb,prl]{revtex4}

\usepackage{amsfonts}

\usepackage{graphicx}
\usepackage{dcolumn}
\usepackage{bm}
\usepackage{times,epsfig,amssymb,amsmath}

\begin{document}

\title{Dirac equation for superluminal neutrinos and mass matrix}

\author{Chang-Yu Zhu$^{1}$\footnote{zhuchangyu2009@163.com}, H. Fan$^{2}$\footnote{hfan@iphy.ac.cn}, and Shi-Ping Ding$^{3}$\footnote{dingsp1985@sina.com}}
\affiliation{%
$^{1}$Department of Physics, Zhengzhou University, Zhengzhou, Henan
450052, China\\
$^{2}$Institute of Physics, Chinese Academy of Sciences, Beijing 100190, China\\
$^{3}$Department of Chemistry, Zhengzhou University, Zhengzhou, Henan
450052, China\\
}%

\date{\today}

\begin{abstract}
The superluminal is reported in OPERA neutrino experiment (arXiv:1109.4897v2), this result is consistent with previous MINOS experiment about neutrino velocity (Phys. Rev. D 76, 072005 (2007)). Here we propose a Dirac equation of tachyon for superluminal neutrino in which the mass matrix is anti-hermitian.
Additionally we find that this anti-hermitian mass matrix which is responsible for superluminal can only be for neutrinos,
while other results of Standard Model are unchanged. In case that Lorentz invariant is remained,
the masses of neutrinos are estimated based on available data of experiments.
\end{abstract}

\maketitle
Recently, it is found in OPERA neutrino experiment that the velocity of muon neutrino can be superluminal.
This result is consistent with previous result of MINOS. The relative difference of the neutrino
velocities with respect to the speed of light are, $v/c-1=(2.37\pm 0.32 ({\rm stat.})^{+0.34}_{-0.24} ({\rm sys.}))\times 10^{-5}$
in OPERA \cite{opera}, and $v/c-1=(5.1\pm 2.9 )\times 10^{-5}$ in MINOS \cite{minos}, respectively,
and possibly in other cases \cite{kamiII,fermi}.
It is apparent that superluminal will cause the change of foundation of modern physics which
is, however, well established. On the other hand, it is also possible that superluminal may really
happen in nature.
So it is necessary to check whether we can have a theory with superluminal neutrinos while still keep
the well-established and confirmed results unchanged. Of course, there are already theories about
tachyon, particle with a superluminal velocity, well before the recent experimental results.
Still we need to propose a theory to fit those new developments.

It is known that we have the mass-energy relation,
\begin{eqnarray}
\left( \frac {v}{c}\right) ^2-1=-c^4\frac {m^2}{E^2},
\label{massenergy}
\end{eqnarray}
where $c$ is the speed of light, $v$ is the measured speed of neutrinos.
According to the experimental results in MINOS and OPERA,
the left hand side of (\ref{massenergy}) are approximately equal for those two
experiments. In OPERA experiment, the CNGS beam is an almost pure $\nu _{\mu }$
beam with an average energy of 17GeV, while in MINOS, still with $\nu _{\mu }$
the average energy is
3GeV. It seems that the right hand side of Eq.(\ref{massenergy}) are different
for two experiments, this is also the case compared with previous experimental
results in SN1987a Ref.\cite{kamiII}. Recently various theories have been
proposed concerning about the possible superluminal results \cite{superlum1}.
In this paper,
we mainly concern about the negative symbol in the right hand side of Eq.(\ref{massenergy}),
and try to explain why the superluminal is only for neutrinos,
we might consider the original mass-energy relation (\ref{massenergy}) itself unchanged
or with possible extensions for superluminal
neutrinos. The discrepancies for right hand side of Eq.(\ref{massenergy}) in experiments
might due to neutrino oscillation, some experiments actually
try to measure the neutrino masses with different mass eigenstates.
While most of the theories are concerning about the Lorentz violation/modification,
in case the Lorentz invariant is remained, with available experimental
data of energy levels and neutrino velocities, here in this paper, we will
provide an estimation of masses of neutrinos.

Our main motivation is that we should let square of mass
$m^2$ itself be negative for neutrinos. Let us consider what is the problem with an imaginary
mass $m^*=-m$, where $*$ means the complex conjugation. The Dirac equation is written as,
\begin{eqnarray}
\left( i\gamma ^{\mu }\partial _{\mu }-m\right)\psi =0,
\end{eqnarray}
where $\gamma ^{\mu }$ are Dirac matrices. We then have the conjugation equation,
\begin{eqnarray}
\overline{\psi }\left( i\gamma ^{\mu }\partial _{\mu }+m\right)=0.
\end{eqnarray}
Note that the sign of mass $m$ then becomes positive since it is imaginary. With standard definition of
current and energy-momentum tensor, $j^{\mu }=\overline{\psi }\gamma ^{\mu }\psi $ and ${T^{\mu }}_{\nu }=
\frac {1}{2}
i\overline{\psi }\left( \gamma ^{\mu }\partial _{\nu }+ \gamma _{\nu }\partial ^{\mu }\right)\psi $, we may
then find the current conservation and energy-momentum conservation laws cannot hold:
$\partial _{\mu }j^{\mu }\not= 0$, $\partial _{\mu }{T^{\mu }}_{\nu }\not =0$.

Next we will try to show that in the representation of the unified weak-electric theory,
there is a symbol for neutrino which can
be either positive or negative in metric. If we choose negative symbol for neutrino, we then can have an anti-hermitian mass matrix
which leads to superluminal. Then we have the current conservation law and energy-momentum conservation law with this modified
metric.

The isospin operators take the form,
\begin{eqnarray}
I_1=\left( \begin{array}{cccc}
0&0&1&0 \\ 0&0&0&0 \\ 1&0&0&0\\0&0&0&0
\end{array}\right),
I_2=\left( \begin{array}{cccc}
0&0&-i&0 \\ 0&0&0&0 \\ i&0&0&0\\0&0&0&0
\end{array}\right),
I_3=\left( \begin{array}{cccc}
1&0&0&0 \\ 0&0&0&0 \\ 0&0&-1&0\\0&0&0&0
\end{array}\right),
\end{eqnarray}
they satisfy the relation $[I_i,I_j]=\frac {1}{2}i\epsilon _{ijk}I_k$, where $\epsilon _{ijk}$ is antisymmetric operator. The hypercharge operator
is,
\begin{eqnarray}
Y=
\left( \begin{array}{cccc}
-1&0&0&0 \\ 0&0&0&0 \\ 0&0&-1&0\\0&0&0&-2
\end{array}\right) .
\end{eqnarray}
It is shown that there are three generations of neutrinos which are $\nu _e, \nu _{\mu },
\nu _{\tau }$. Each generation of leptons may be represented by four bases distinguished by eigenvalues of isospin operator $I_3$, $0,\frac {1}{2},-\frac {1}{2}$ and hypercharge operator eigenvalues of $Y$, $-1,0,-1,-2$. Explicitly, the four eigenstates are $|Y,I_3\rangle $
which take the forms: $|-1,\frac {1}{2}\rangle $, $|0,0\rangle $,
$|-1,-\frac {1}{2}\rangle $, $|-2,0\rangle $, where the first two are for
left and right neutrinos, the last two are for left and right electrons.

Now we propose a new metric for superluminal neutrino as,
\begin{eqnarray}
h=\left( \begin{array}{cccc}
1&0&0&0 \\ 0&-1&0&0 \\ 0&0&1&0\\0&0&0&1
\end{array}\right) .
\label{metric}
\end{eqnarray}
The difference from standard definition is only that the second diagonal element has a negative symbol.
Still the commutation relations, $[h,Y]=0$, $[h,I_i]=0,i=1,2,3$, remain unchanged which
are explicit with their matrix representations. With the metric (\ref{metric}), now it seems naturally to
have a mass matrix like the following,
\begin{eqnarray}
\hat {m}=\left( \begin{array}{cccc}
0&im_{\nu _{\mu }}&0&0 \\ im_{\nu _{\mu }}&0&0&0 \\ 0&0&0&m_{\mu }\\0&0&m_{\mu }&0
\end{array}\right) ,
\label{mass}
\end{eqnarray}
where $im_{\nu _{\mu }}$ is the muon neutrino mass which is an imaginary number
($m_{\nu _{\mu }}$ is positive),
$m_{\mu }$ is the corresponding mass of electric lepton.
Now $\hat {m}$ and $h$ do not commute. With this mass matrix, we then have a negative square
of mass for muon neutrino but with the corresponding mass for electric lepton unchanged.
Explicitly for (\ref{mass}), we have the diagonal elements of
$\hat {m}^2$ as $(-m_{\nu _{\mu }}^2, -m_{\nu _{\mu }}^2,m_{\mu }^2, m_{\mu }^2)$.

As in Standard Model, we now define, $t_1=\frac {1}{2}g_1Y$, $t_2=g_2I_1,t_3=g_2I_2,t_4=g_2I_3$, where $U(1)$ gauge coupling
takes the value, $g_1\approx 0.36$,
$SU(2)$ gauge coupling is, $g_2\approx 0.65$. The weak-electric gauge field takes the form,
$\hat {A}_{\mu }=A_{\mu }^at_a$. The Dirac equation is written as,
\begin{eqnarray}
\left(i\gamma ^{\mu }\partial _{\mu }+\gamma ^{\mu }A_{\mu }^at_a-\hat {m}\right)\psi =0.
\end{eqnarray}
By metric $h$ in (\ref{metric}), we can now define $\overline{\psi }=\psi ^{\dagger }\gamma _0h$,
and for mass matrix we have
\begin{eqnarray}
\overline{\hat {m}}=h^{-1}\hat {m}^{\dagger }h=\hat {m},
\end{eqnarray}
here negative signs in $\hat {m}^{\dagger }$ are complemented
by the negative symbols in $h$. The original equations $\overline{t_a}=h^{-1}t_a^{\dagger }h=t_a$ still hold.
So the dual Dirac equation now is written,
\begin{eqnarray}
\overline{\psi }\left( i\gamma ^{\mu }\partial _{\mu }+
\gamma ^{\mu }A^a_{\mu }t_a-\hat {m}\right) =0.
\end{eqnarray}
With all those results, we then have the current and the energy-momentum tensor,
\begin{eqnarray}
j^{\mu }&=&\overline{\psi }\gamma ^{\mu }\psi ,\\
{T^{\mu }}_{\nu }&=&\frac{1}{2}
i\overline{\psi }\left( \gamma ^{\mu }\partial _{\nu }+ \gamma _{\nu }\partial ^{\mu }\right)\psi .
\end{eqnarray}
Now the current conservation and the energy-momentum conservation law are satisfied,
\begin{eqnarray}
\partial _{\mu }j^{\mu }&=& 0,\\
 \partial _{\mu }{T^{\mu }}_{\nu } &=&0.
 \end{eqnarray}

The other two generations of neutrinos can be similarly studied. In all, the mass matrix with
all three generations of neutrinos can be written as three times three matrix, square
of masses should be negative or zero measured with orthogonal mass eigenstates.
As in (\ref{mass}), $m_{\nu _{\mu }}$ is now replaced by a three times three mass matrix which
itself is hermitian while the additional imaginary unit $i$ is presented before it. Under
a unitary transformation, this matrix can be diagonalized as $(m_1,m_2,m_3)$ which is
semi-definite. The corresponding metric takes a diagonal form with non-zero elements as $(I,-I,I,I)$,
where $I$ is a three-level identity matrix, some notations may be found, for example in Ref.\cite{zf}.

Since enough experimental data for mass-matrix of the neutrinos is still absent, it is not
clear what is the exact form of it. However,
we show that superluminal neutrinos with square of the mass being negative
can be fitted into the original Standard Model without changing other results. Also we might
provide an estimation of this mass matrix.
Though a lot of recent
theories consider a violation/modification of Lorentz invariant under various assumputions/circumstances, here let us consider the consequences that the Lorentz invariant is unchanged.  In any experiment, the measured mass is actually an eigenvalue of the mass matrix
with a fixed eigenvector. Those eigenvalues can be calculated as :
120MeV in Ref.\cite{opera}; 30MeV in Ref.\cite{minos}; 0.32KeV in Ref.\cite{kamiII};
0.27GeV in Ref.\cite{fermi}, note that the imaginary unit is already available in (\ref{mass}).
Thus we estimate that some kind of neutrinos can be quite heavy, whose mass is hundreds times of
electron mass, while some kind of neutrinos can be quite light which is only one over several
hundreds of an electron mass. Here bear in mind that the mass of the neutrino is an imaginary
value, thus there is no energy threshold to create neutrinos. This might be a reason that
neutrino is generally assumed to be quite light. Here in this paper, if we consider that the Lorentz
invariant is remained, masses of different neutrinos might range from hundreds times
to one percent of an electron mass.

\end{document}